\documentclass[preprintnumbers,10pt,nofootinbib]{revtex4}
\usepackage{amsmath,latexsym,amssymb,amsfonts}
\usepackage{epsfig}
\usepackage{bm}
\usepackage{float}
\usepackage{xcolor}

\begin{document}

\title{\textbf{Before the Page time: maximum entanglements or the return of the monster?}}

\author{
{\textsc{Jeong-Myeong Bae$^{a}$}}\footnote{{\tt bjmhk2{}@{}dgist.ac.kr}},
{\textsc{Dong Jin Lee$^{b}$}}\footnote{{\tt dj0626{}@{}kaist.ac.kr}},
{\textsc{Dong-han Yeom$^{c,d}$}}\footnote{{\tt innocent.yeom{}@{}gmail.com}}
and 
{\textsc{Heeseung Zoe$^{a}$}}\footnote{{\tt heezoe{}@{}dgist.ac.kr}}
}

\affiliation{
$^{a}$\small{School of Undergraduate Studies, College of Transdisciplinary Studies,\\  
Daegu Gyeongbuk Institute of Science and Technology (DGIST), Daegu 42988, Republic of Korea}\\
$^{b}$\small{Department of Physics, KAIST, Daejeon 34141, Republic of Korea}\\
$^{c}$\small{Department of Physics Education, Pusan National University, Busan 46241, Republic of Korea}\\
$^{d}$\small{Research Center for Dielectric and Advanced Matter Physics, Pusan National University, Busan 46241, Republic of Korea}
}

\begin{abstract}
The conservation of information of evaporating black holes is a very natural consequence of unitarity which is the fundamental symmetry of quantum mechanics. In order to study the conservation of information, we need to understand the nature of the entanglement entropy. The entropy of Hawking radiation is approximately equal to the maximum of entanglement entropy if a black hole is in a state before the Page time, i.e., when the entropy of Hawking radiation is smaller than the entropy of the black hole. However, if there exists a process generating smaller entanglements rather than maximal entanglements, the entropy of Hawking radiation will become smaller than the maximum of the entanglement entropy before the Page time. If this process accumulates, even though the probability is small, the emitted radiation can eventually be distinguished from the exactly thermal state. In this paper, we provide several interpretations of this phenomenon: (1) information of the collapsed matter is emitted before the Page time, (2) there exists a firewall or a non-local effect before the Page time, or (3) the statistical entropy is greater than the areal entropy; a monster is formed. Our conclusion will help resolve the information loss paradox by providing groundwork for further research.
\end{abstract}

\maketitle

\newpage

\tableofcontents

\section{Introduction}

The information loss paradox is an unresolved problem that contains the essential nature of the quantum theory of gravity \cite{Hawking:1976ra}. If we assume semi-classical general relativity and the unitarity of quantum mechanics, then inconsistent conclusions follow \cite{Yeom:2008qw}. Accordingly, we must further clarify the five assumptions as follows \cite{Susskind:1993if}:
\begin{itemize}
\item[-- A1.] \textit{Unitarity}: There is no loss of information from black hole formation to complete evaporation.
\item[-- A2.] \textit{Local quantum field theory}: The computation of Hawking radiation is consistent.
\item[-- A3.] \textit{General relativity}: Classical description is sufficient unless the curvature is larger than the Planck scale.
\item[-- A4.] \textit{Bekenstein-Hawking entropy as statistical entropy}: The Boltzmann entropy carried by a black hole is expressed by $A/4$, where $A$ denotes the horizon area.
\item[-- A5.] \textit{The existence of an information observer}: There exists an observer who can read and distinguish information from Hawking radiation.
\end{itemize}
These five assumptions cannot be consistent. In essence, either there exists an observer who can detect the duplication of information \cite{Yeom:2008qw} or the laws of entanglement entropy are violated \cite{Almheiri:2012rt}.

The majority of candidate resolutions of the information loss paradox can be categorized by denying at least one of the five assumptions \cite{Sasaki:2014spa}. These are outlined below:
\begin{itemize}
\item[-- R1.] Violation of \textit{unitarity}: Information can be lost \cite{Hawking:1976ra,Unruh:1995gn}.
\item[-- R2.] Violation of the \textit{local quantum field theory}: non-local interactions exist between inside and outside the horizon \cite{Giddings:2012gc}.
\item[-- R3.] Violation of \textit{general relativity}: There may exist a membrane \cite{Thorne:1986iy}, a fuzzball \cite{Mathur:2005zp}, or a firewall \cite{Almheiri:2012rt} approximal to the horizon.
\item[-- R4.] Existence of \textit{monsters}: Information can be retained by a small or large remnant \cite{Chen:2014jwq}. Such a remnant can be a type of baby universe \cite{Hwang:2010gc}.
\item[-- R5.] Absence of an \textit{information observer}: No (semi-classical) observer exists to read information from Hawking radiation, and therefore, information is effectively lost \cite{Lee:2015rwa}.
\end{itemize}

In our discussion, our arguments strongly rely on the result of the Page curve \cite{Page:1993df}. According to \cite{Page:1993df}, when we consider a pure and random bipartite system ($n$ and $m$ are the number of states for each subsystem and $1 \ll m \leq n$), the entanglement entropy, $S$, between two subsystems is approximated
\begin{eqnarray}
S \simeq \log m - \frac{m}{2n}.
\end{eqnarray}
Therefore, small subsystems are always approximately in the thermal state because the entanglement entropy is essentially the same as the Boltzmann entropy \cite{Lloyd:1988cn}, i.e.,
\begin{eqnarray}
\log m - S \ll 1.
\end{eqnarray}
This indicates that we cannot obtain distinguishable information from Hawking radiation if the number of radiation states is sufficiently smaller than states of inside the black hole. If we apply this to an evaporating black hole, then information can only be obtained after the \textit{Page time}, i.e., the time when two subsystems have the same number of states ($m = n$). After the Page time, the number of states outside the horizon becomes larger than the number inside, and therefore, it is possible to read the information outside the horizon. Accordingly, it is not surprising that many contradictory situations occur after the Page time \cite{Yeom:2008qw,Almheiri:2012rt}, which is the main focus of many papers.

\begin{figure}
\begin{center}
\includegraphics[scale=0.45]{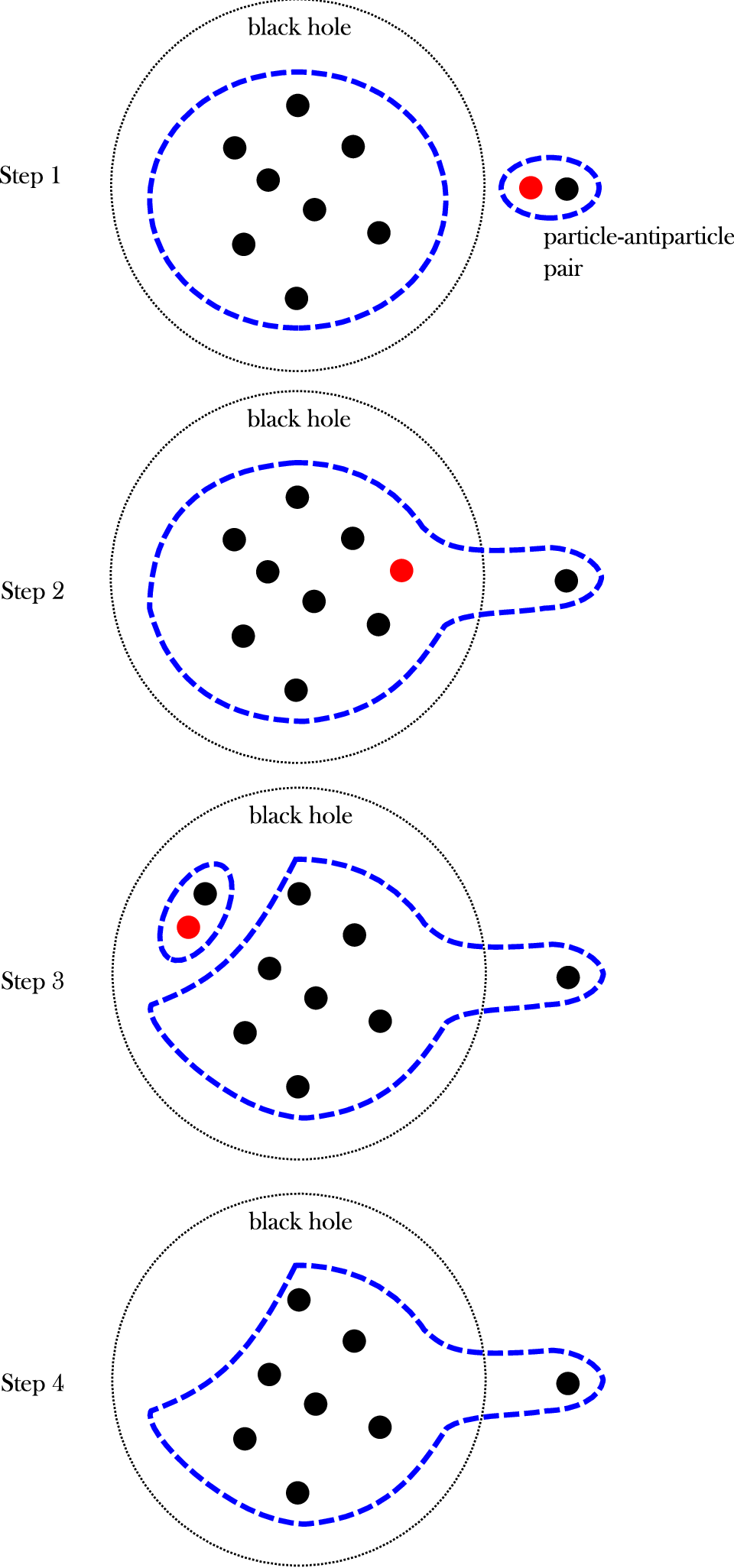}
\caption{\label{fig:Fig}Hawking radiation before the Page time. Blue dashed curves denote the boundary of entanglements. Step 1: Due to assumptions A1 and A3, a particle-antiparticle pair is generated as a separable state. Step 2: The antiparticle carrying negative energy falls into the horizon. Step 3 and Step 4: Due to assumptions A1 and A4, the negative energy antiparticle should be unitarily annihilated (Credit: \cite{Hwang:2017yxp}).}
\end{center}
\end{figure}

Contrastingly, in this paper, we examine events \textit{before} the Page time. Before the Page time, the number of states inside a black hole is dominated. As Hawking particles are emitted, the entanglements between the inside and outside of a black hole linearly increase \cite{Hwang:2016otg}. Furthermore, the increasing entanglement entropy takes on a maximum value and, accordingly, the thermal nature is preserved. The problem remains, however, of whether this phenomenon is a natural consequence. The assumptions for this are outlined below (Fig.~\ref{fig:Fig}) \cite{Hwang:2017yxp}:
\begin{itemize}
\item[-- A1/A3]: A pair-creation of particles is a unitary process near the horizon and the background is nothing special near the horizon. Hence, the pair-creation process is described by the \textit{creation of a separable state}\footnote{In this paper, we consider the creation of a particle-antiparticle pair (or the combination of many particles as a separable state), where separable state refers to no entanglement between the previous particles (black hole or radiation) and the created particles; however, of course, there must exist entanglements among the created particles.}.
\item[-- A2]: All interactions are local. Thus, after the antiparticles fall into the black hole, they only interact with the degrees of freedom inside the black hole (and vice versa). Since there are no interactions between inside and outside the horizon, to preserve maximum entanglements (following the Page argument), the Hawking particles themselves must be \textit{maximally entangled}.
\item[-- A1/A4]: After particle emission, the black hole area decreases. Hence, there should exist a unitary annihilation process between antiparticles and the internal degrees of freedom. In other words, there exists a process consisting of the \textit{annihilation of a separable state}.
\end{itemize}
If these three key points are repeated, then one can explain the emission of particles, at least, before the Page time \cite{Hwang:2017yxp}.

In this paper, we examine whether these three key points are consistent with each other. If Hawking radiation is dominated by free fields, then it is not surprising that Hawking radiation can be described by maximally entangled pairs. However, if the field is interacting, processes more complicated than a single particle and single antiparticle processes can exist. In addition, even though a particle-antiparticle pair is created with maximum entanglements, if two pairs interact with each other, then it is possible for the maximum entanglement assumption to be violated; accordingly, the two particles and two antiparticles are totally randomized. In turn, this raises the problem of whether the Page curve is valid before the Page time. By using spin-1/2 toy models \cite{Hwang:2016otg,Hwang:2017yxp,Bae:2019niw}, we demonstrate the possibility of this alongside various interpretations.

This paper is organized as follows. In Sec.~\ref{sec:pre}, we discuss several preliminary topics; in Sec.~\ref{sec:bia}, we examine whether the field is interacting, from which the bias from the maximum entanglements can be accumulated; in Sec.~\ref{sec:pos}, we discuss the potential implications of our conclusion; and in Sec.~\ref{sec:con}, we summarize the paper and discuss possible future research directions.

\section{\label{sec:pre}Preliminaries}

\subsection{Entropy increase due to Hawking radiation}

First, let us consider a situation that the Hawking process creates a particle-antiparticle pair. If we only impose the separable state condition, then maximum entanglements are not necessarily created. For example, if we consider spin-1/2 particles, then there are four bases of the separable combinations:
\begin{eqnarray}
\left| \ell = 1,\;\;  m = 1 \rangle\right. &=& \left| \uparrow  \rangle\right. \left| \uparrow  \rangle\right.,\\
\left| \ell = 1,\;\;  m = 0 \rangle\right. &=& \frac{1}{\sqrt{2}} \left( \left| \uparrow  \rangle\right. \left| \downarrow  \rangle\right. + \left| \downarrow  \rangle\right. \left| \uparrow  \rangle\right. \right),\\
\left| \ell = 1,\;\;  m = -1 \rangle\right. &=& \left| \downarrow  \rangle\right. \left| \downarrow  \rangle\right.,\\
\left| \ell = 0,\;\;  m = 0 \rangle\right. &=& \frac{1}{\sqrt{2}} \left( \left| \uparrow  \rangle\right. \left| \downarrow  \rangle\right. - \left| \downarrow  \rangle\right. \left| \uparrow  \rangle\right. \right),
\end{eqnarray}
where $\ell$ and $m$ denote the quantum numbers of total states. The second and fourth bases are maximally entangled states. This implies that if we choose a \textit{random} particle-antiparticle pair, then maximum entanglements are not guaranteed.

Let us assume that there is a bipartite system, where the subsystems are $A$ (black hole) and $B$ (radiation). For this given $A \cup B$ system, a particle-antiparticle pair is created, acting as a separable and random state ($c_{1} \cup c_{2}$); here, one ($c_{1}$) falls into the horizon and the other ($c_{2}$) remains outside. By assuming locality, $A$ only interacts with $c_{1}$ and $B$ only interacts with $c_{2}$. Assuming $A$ and $B$ are initially in the random state, we can examine the behavior of the entanglement entropy after $c_{1}$ and $c_{2}$ are created by using numerical simulations with spin-1/2 systems, as shown in Fig.~\ref{fig:sim3} (for mathematical details, see Appendix A) \cite{Hwang:2017yxp}.

\begin{figure}
\begin{center}
\includegraphics*[scale=0.5,viewport=50 210 500 600]{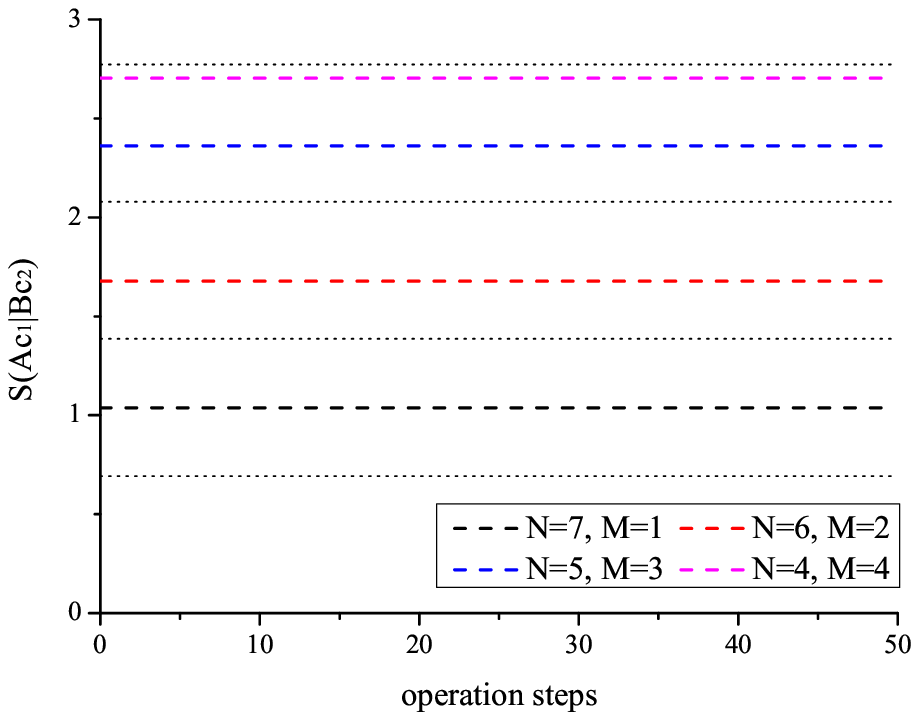}
\includegraphics*[scale=0.5,viewport=50 210 500 600]{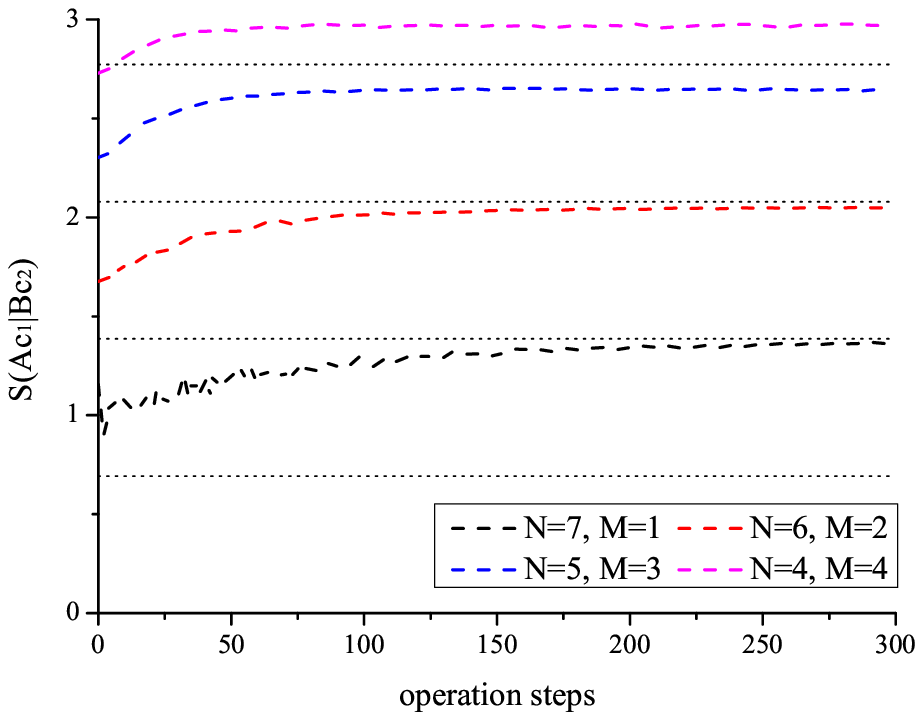}
\caption{\label{fig:sim3}Left: $S(A\cup c_{1}|B\cup c_{2})$, where the interaction is only local. Right: $S(A\cup c_{1}|B\cup c_{2})$, where the interaction is non-local. Horizontal lines are $\log 2$, $2 \log 2$, $3 \log 2$, and $4 \log 2$ from bottom to top (Credit: \cite{Hwang:2017yxp}).}
\end{center}
\end{figure}

Let us assume that a black hole is in a state before the Page time and that $N$ and $M$ denote the number of particles inside and outside the black hole, respectively. For this combination, the Page limit of the entropy is $S = \log m = M \log 2$. Before we add the particles, the initial state approximately satisfies this limit (unless $N \sim M$). However, after we add $c_{1}$ and $c_{2}$, the entanglement entropy slightly increases (left of Fig.~\ref{fig:sim3}), which can be attributed to the entanglement between $c_{1}$ and $c_{2}$; however, since $c_{1}$ and $c_{2}$ are not maximally entangled, the system is not within the Page limit. In addition, even though we turn on the interactions between $A \cup c_{1}$ and $B \cup c_{2}$ (as an example, we used the swap operation; see Appendix B) the entanglement entropy between $A\cup c_{1}$ and $B \cup c_{2}$ does not change. Alternatively, if we turn on interactions between $A\cup c_{1}$ and $B \cup c_{2}$ (non-local interactions between inside and outside the horizon), the said entropy increases and approaches the Page limit (right of Fig.~\ref{fig:sim3})\footnote{Note that operation steps may indicate the direction of time, but we do not specify the detailed time-dependence.}.

This conclusion is not surprising because the Page limit condition ($\log m - S \ll 1$) implies a random state. After we add $c_{1}$ and $c_{2}$, the total system cannot be random, unless, of course, $c_{1}$ and $c_{2}$ are sufficiently random from the beginning (i.e., maximal entanglements) or unless there is a sufficient additional randomization process between the inside and outside (i.e., non-local interactions). Accordingly, to ensure the Page limit and the maximum entanglement condition take place before the Page time, either \textit{the Hawking process is always concerned with maximum entanglements} or \textit{there exists non-local interactions}.

\subsection{Entanglements from particle pair-productions}

\begin{figure}
\begin{center}
\includegraphics[scale=0.75]{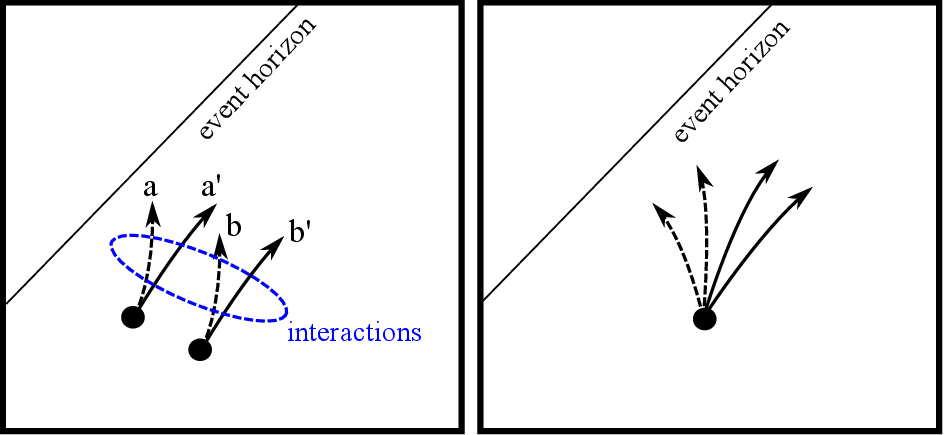}
\caption{\label{fig:inter}Left: Two pairs are maximally entangled, but these four particles can interact outside the horizon. Right: Four particles are created from the vacuum and there is no reason to assume that two antiparticles are in the maximum entanglements with two particles.}
%\end{center}
%\end{figure}
%\begin{figure}
%\begin{center}
\includegraphics[scale=0.4]{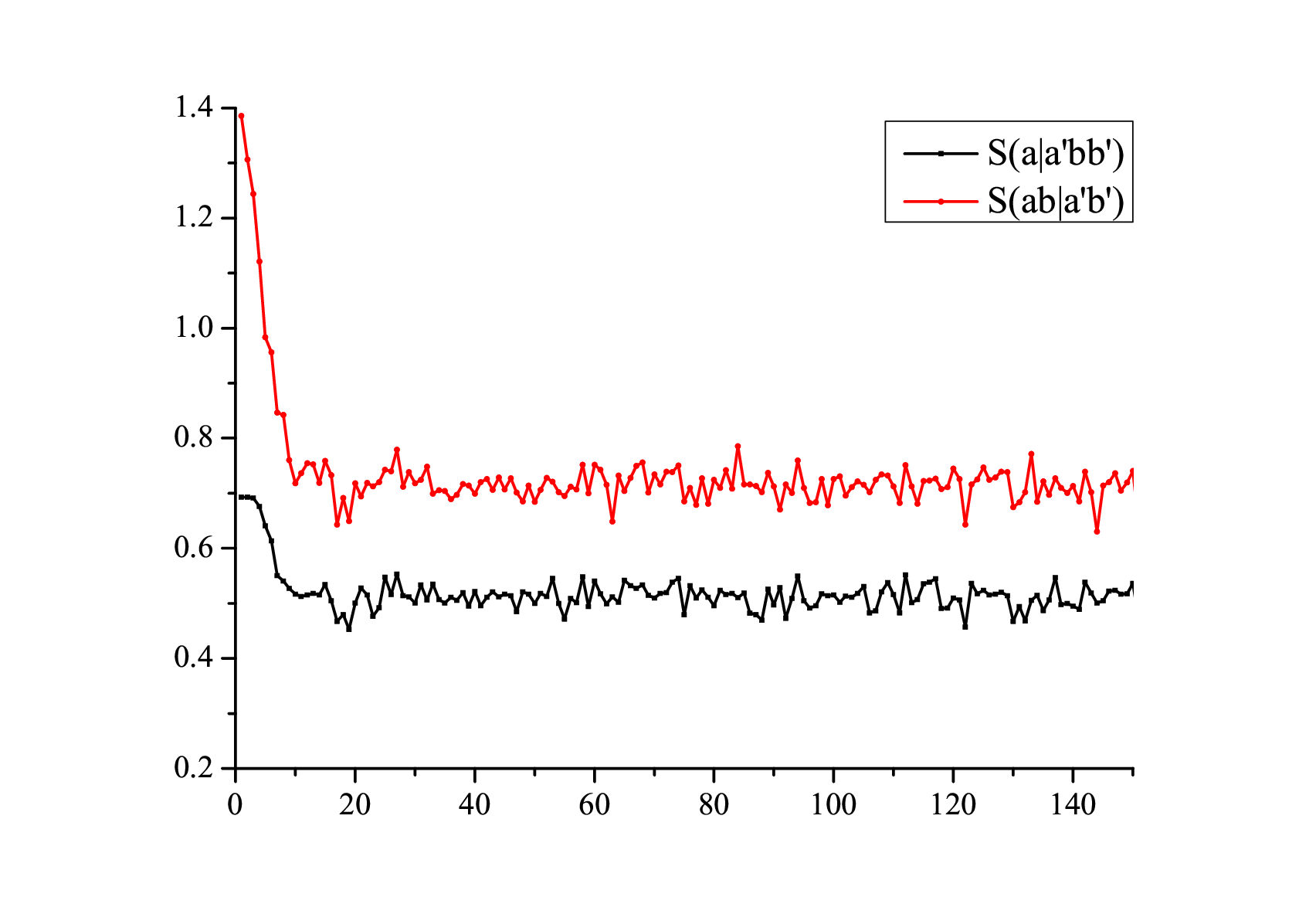}
\caption{\label{fig:randomizing}Randomization of two maximally entangled particle pairs, where the horizontal axis is the number of interactions.}
\end{center}
\end{figure}

In this section, we show that the Hawking process can include particle creations that are not maximally entangled. If the process is dominated by a free field, then we have a good reason to believe that the said process is dominated by maximally entangled pairs because we can explicitly construct the Israel-Hartle-Hawking state \cite{Israel:1976ur}. However, if the fields can interact with each other, then the created particles may not carry the maximal entanglements (Fig.~\ref{fig:inter}).

For example, if two maximally entangled pairs can communicate outside the event horizon, then four particles are randomized. Hence, two incoming particles and two outgoing particles cannot preserve the maximum entanglements (left of Fig.~\ref{fig:inter}). We imagine that there are two maximally entangled pairs ($a$-$a'$ and $b$-$b'$), the four particles of which interact via swap operations as time goes on (Fig.~\ref{fig:randomizing}). As we calculate the entanglement entropy of $a$ or $a \cup b$, from the beginning, the entropy is maximal ($\log 2$ and $2 \log 2$, respectively), but as the number of interactions increases, the entropy quickly decreases\footnote{Note that from the maximal entanglement, if one only uses the swap operations, the final state may not be a totally randomized state because the swap operation does not change the numerical value of the density matrix; rather, the operation only switches each number. However, in realistic interactions, it is reasonable to assume that the numerical values can be changed; hence, it is reasonable to conclude that the generic interactions will quickly and completely randomize maximally entangled pairs.}. Therefore, if an interaction exists between particles, it is reasonable to conclude that the entanglement entropy between incoming and outgoing particles is less than the maximum value. Moreover, if quantum field theory allows four-particle creation from the vacuum, then there is no reason to think that two incoming particles must be in maximum entanglement with the two outgoing particles (right of Fig.~\ref{fig:inter}). Unfortunately, it is beyond the scope of this paper to construct an exact model for this; however, although such a phenomenon can be probabilistically suppressed, it is reasonable to assume a process where in a vacuum creates randomly mixed states that are not maximum entanglements. If one can accept this possibility, then the entanglement entropy contribution of the half of the randomized four-particle system must be estimated, as well as the six-particle system, eight-particle system, and so on. Moreover, if the number of particles is too small, then the Page's approximated result might not be exact and, accordingly, numerical evaluations are required to obtain accurate computations.

\begin{figure}
\begin{center}	
\includegraphics[scale=0.25]{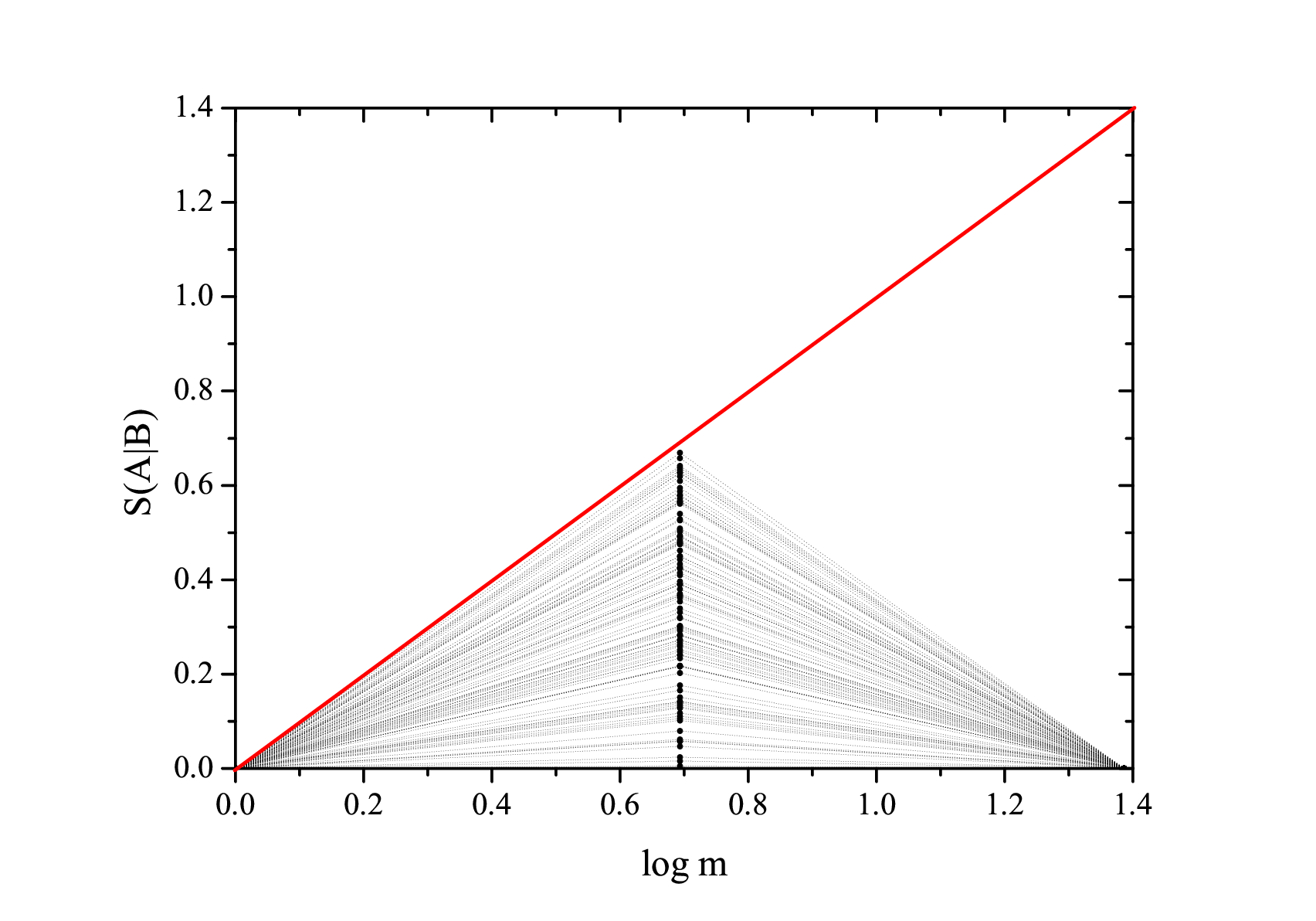}
\includegraphics[scale=0.25]{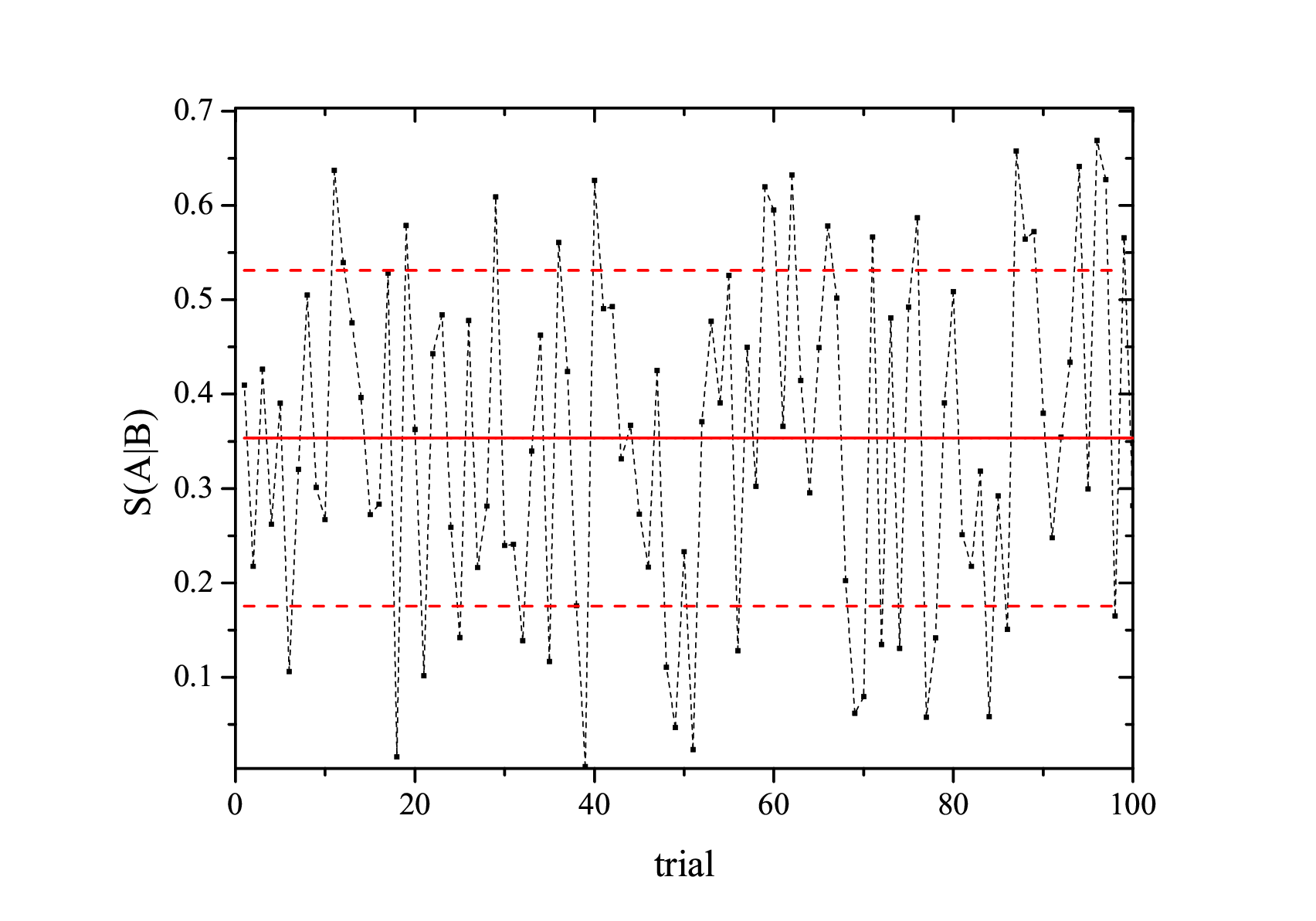}
\includegraphics[scale=0.25]{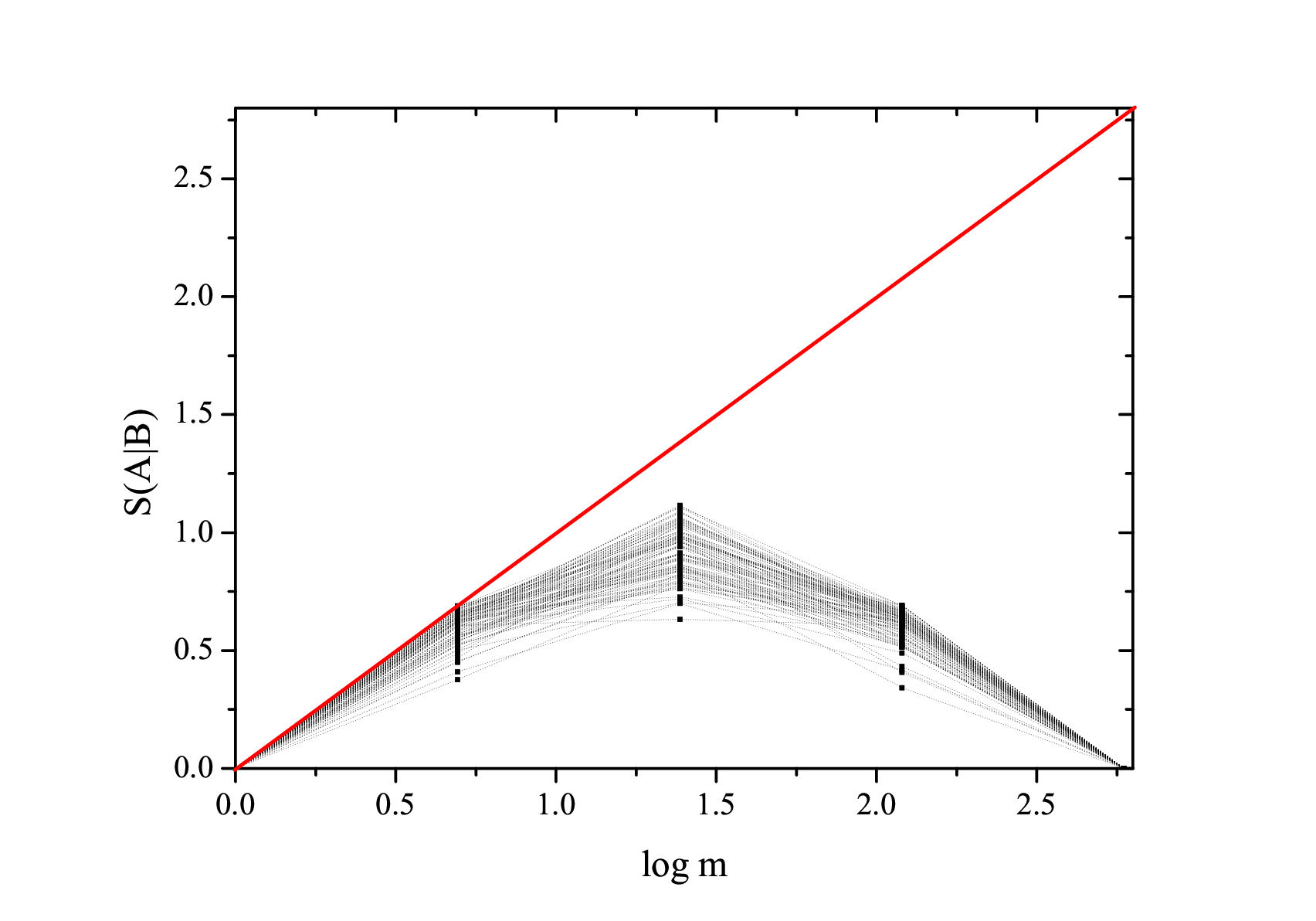}
\includegraphics[scale=0.25]{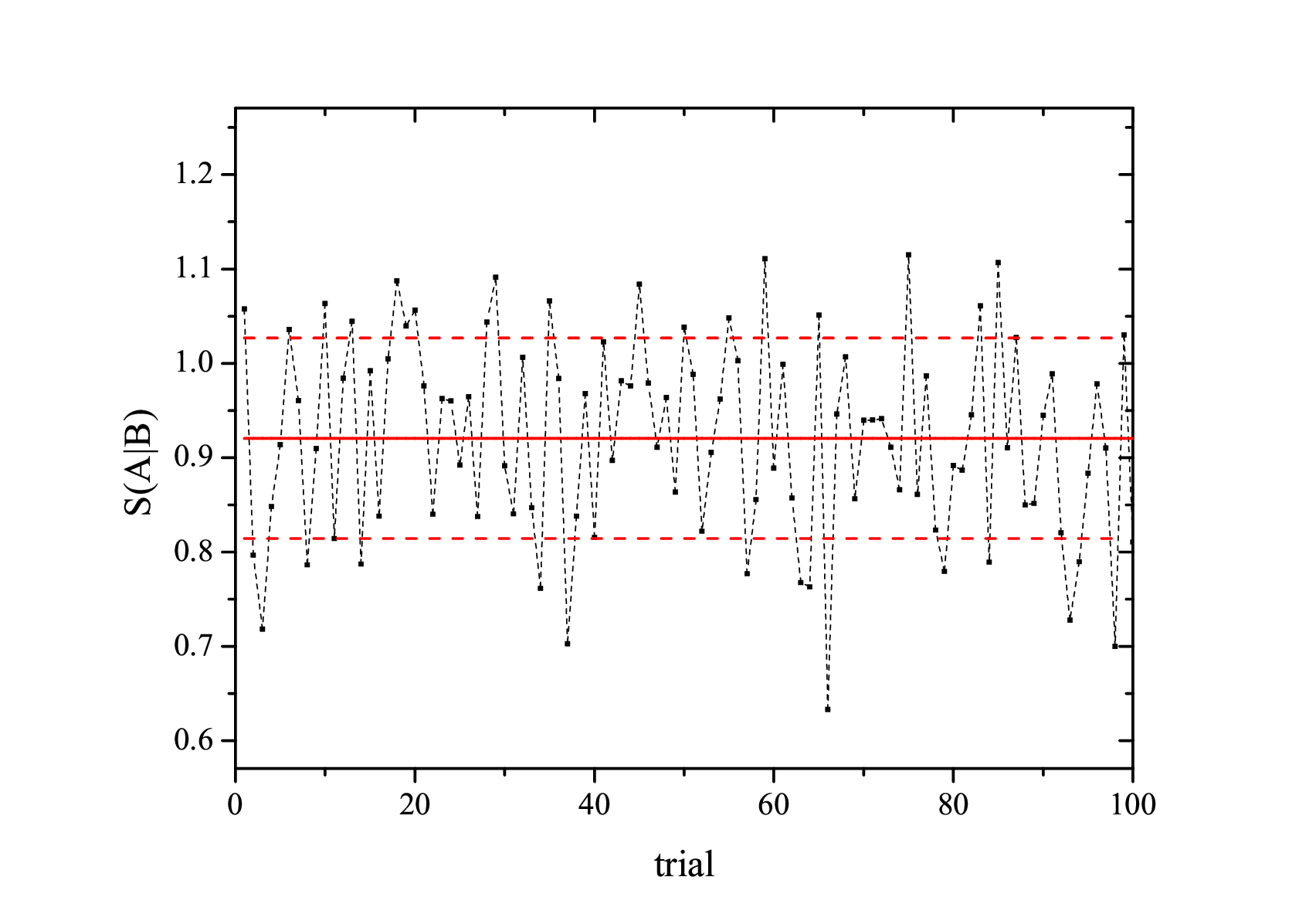}
\includegraphics[scale=0.25]{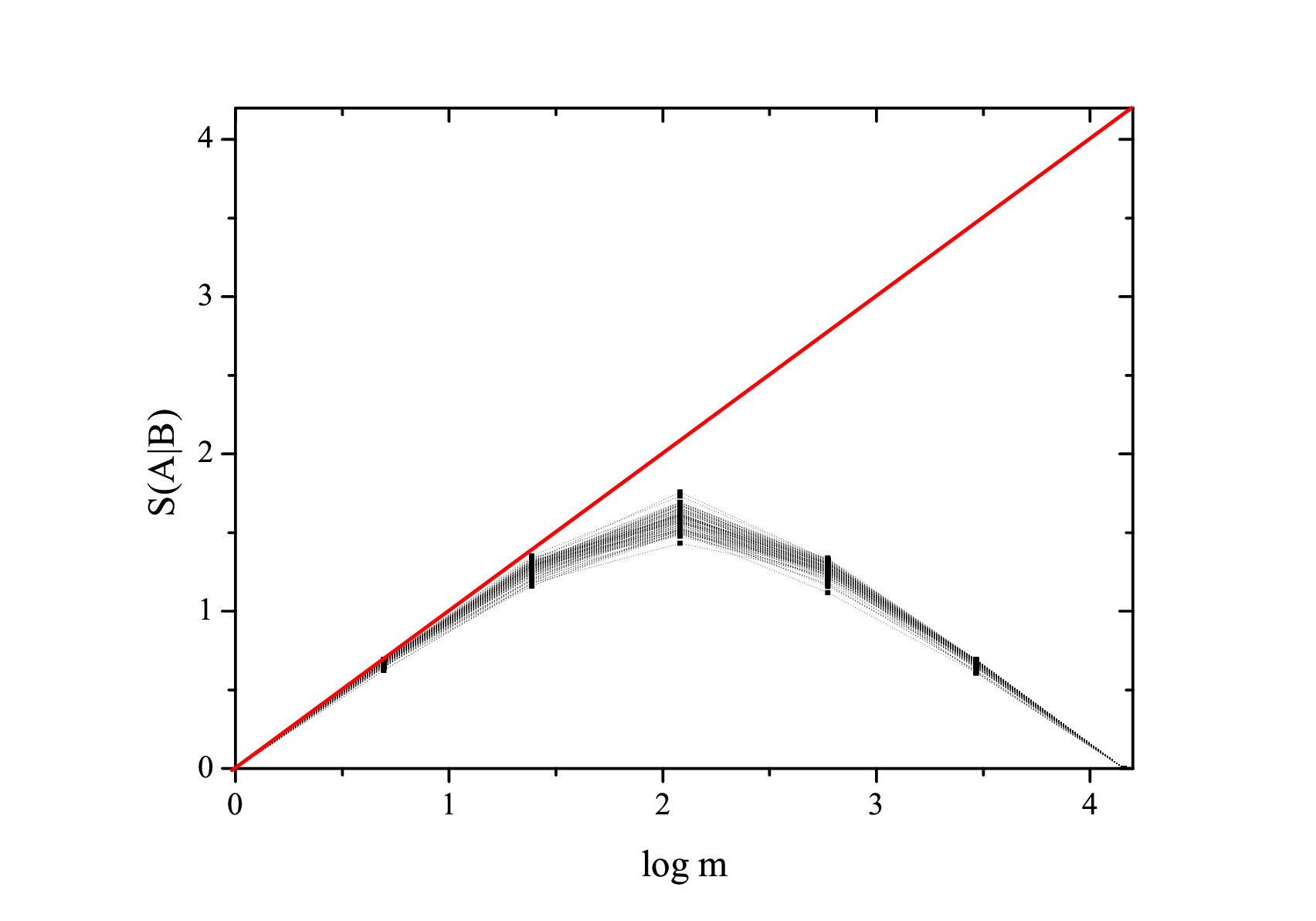}
\includegraphics[scale=0.25]{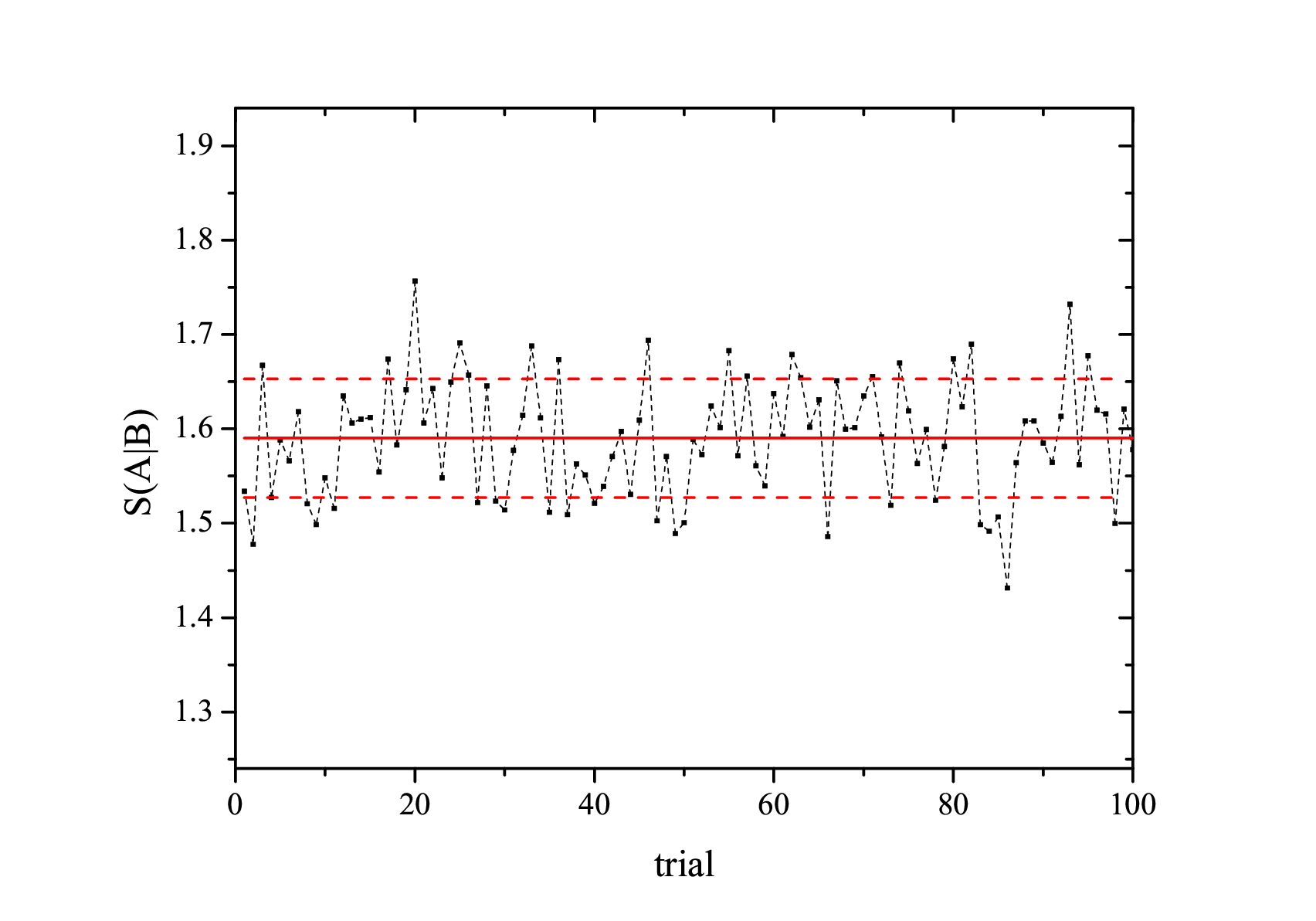}
\includegraphics[scale=0.25]{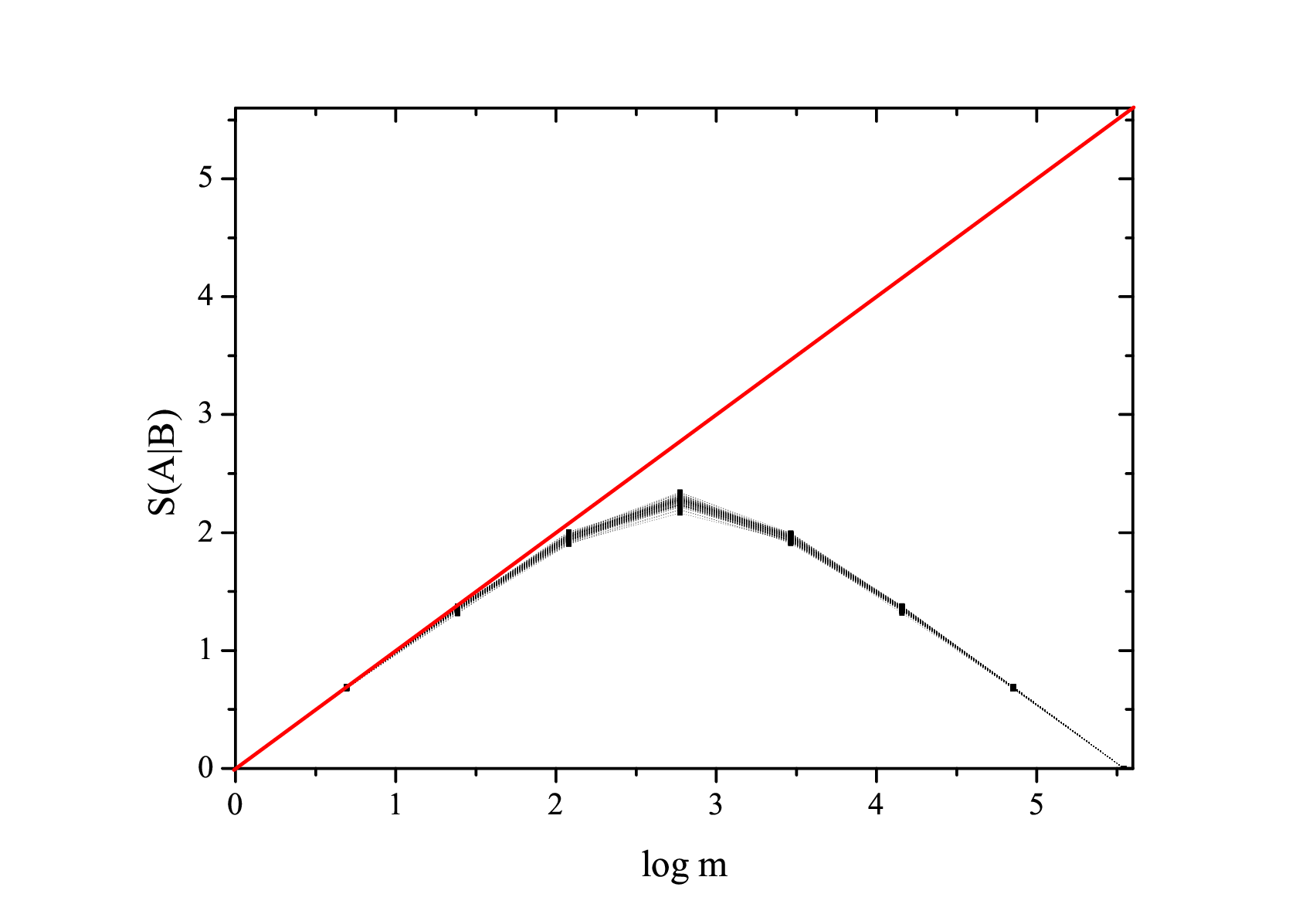}
\includegraphics[scale=0.25]{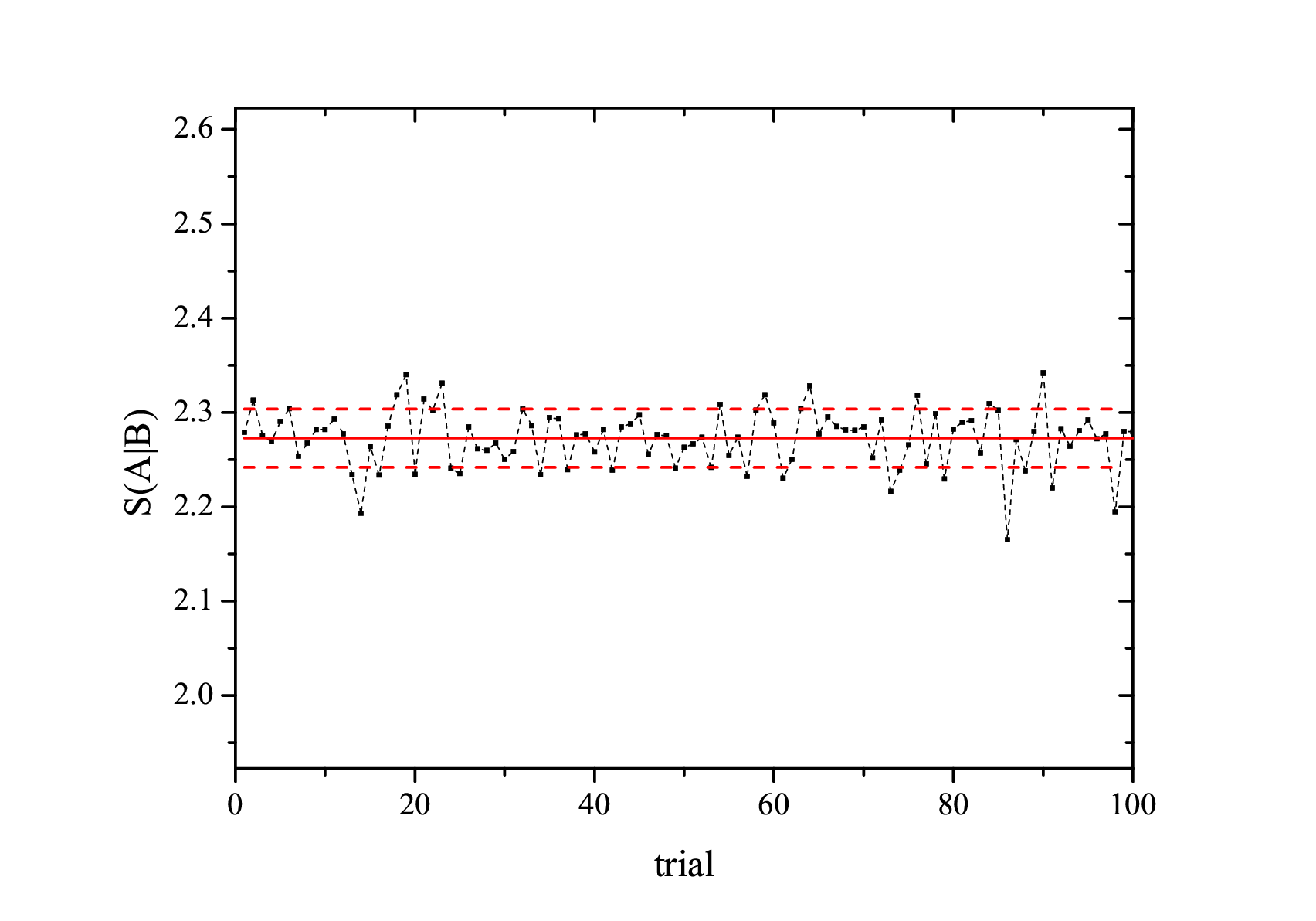}
\caption{Left: Several trials of the Page curve for the two-particle (first), four-particle (second), six-particle (third), and eight-particle (fourth line) systems. Right: Entanglement entropy for the half of the two-particle (first), four-particle (second), six-particle (third), and eight-particle (fourth line) systems, where the red solid line indicates the average of entanglement entropy and the red dashed lines indicate the standard deviations.}
\label{fig:pagecurvedeviation}
\end{center}
\end{figure}

\section{\label{sec:bia}Bias from the Page curve}

We used numerical simulations to estimate the entanglement entropy of the half of a randomly mixed two-, four-, six-, or eight-particle system, the results of which are shown in Fig.~\ref{fig:pagecurvedeviation}. The left figures of Fig.~\ref{fig:pagecurvedeviation} consist of the entanglement entropy curves for a two-, four-, six-, and eight-particle system. By repeating the numerical simulations, many curves were superposed. For each trial, we evaluated the half-way entanglement entropy for a two-, four-, six-, and eight-particle system\footnote{Therefore, the horizontal axis of right figures do not mean the time direction; these mean trials in order to evaluate statistical tendencies.}. This was shown in the right figures of Fig.~\ref{fig:pagecurvedeviation}, from which it is evident that an average value $\langle S(A|B) \rangle$ exists and the result is diversely distributed with a given standard deviation $\sigma$. The numerical results are summarized in Table~\ref{tab}.

\begin{table}[h]
\begin{center}
\begin{tabular}{c|c|c}
\hline \hline
\begin{tabular}{c}
\;\;\;\;\;\;Number of particles\;\;\;\;\;\;
\end{tabular} &
\begin{tabular}{c}
\;\;\;\;\;\;Averaged half-way entanglement entropy\;\;\;\;\;\;\\
$\langle S(A|B) \rangle$
\end{tabular} &
\begin{tabular}{c}
\;\;\;\;\;\;Standard deviation\;\;\;\;\;\;\\
$\sigma$
\end{tabular}\\
\hline \hline
\begin{tabular}{c}
\;\;\;\;\;\;2\;\;\;\;\;\;
\end{tabular} &
\begin{tabular}{c}
\;\;\;\;\;\;0.3533\;\;\;\;\;\;
\end{tabular} &
\begin{tabular}{c}
\;\;\;\;\;\;0.1781\;\;\;\;\;\;
\end{tabular}\\
\hline
\begin{tabular}{c}
\;\;\;\;\;\;4\;\;\;\;\;\;
\end{tabular} &
\begin{tabular}{c}
\;\;\;\;\;\;0.9206\;\;\;\;\;\;
\end{tabular} &
\begin{tabular}{c}
\;\;\;\;\;\;0.1062\;\;\;\;\;\;
\end{tabular}\\
\hline
\begin{tabular}{c}
\;\;\;\;\;\;6\;\;\;\;\;\;
\end{tabular} &
\begin{tabular}{c}
\;\;\;\;\;\;1.5901\;\;\;\;\;\;
\end{tabular} &
\begin{tabular}{c}
\;\;\;\;\;\;0.0628\;\;\;\;\;\;
\end{tabular}\\
\hline
\begin{tabular}{c}
\;\;\;\;\;\;8\;\;\;\;\;\;
\end{tabular} &
\begin{tabular}{c}
\;\;\;\;\;\;2.2727\;\;\;\;\;\;
\end{tabular} &
\begin{tabular}{c}
\;\;\;\;\;\;0.0309\;\;\;\;\;\;
\end{tabular}
\\
\hline \hline
\end{tabular}
\caption{\label{tab}The half-way entanglement entropy $\langle S(A|B) \rangle$ and the standard deviation $\sigma$.}
\end{center}
\end{table}

\begin{figure}
\begin{center}
\includegraphics[scale=0.27]{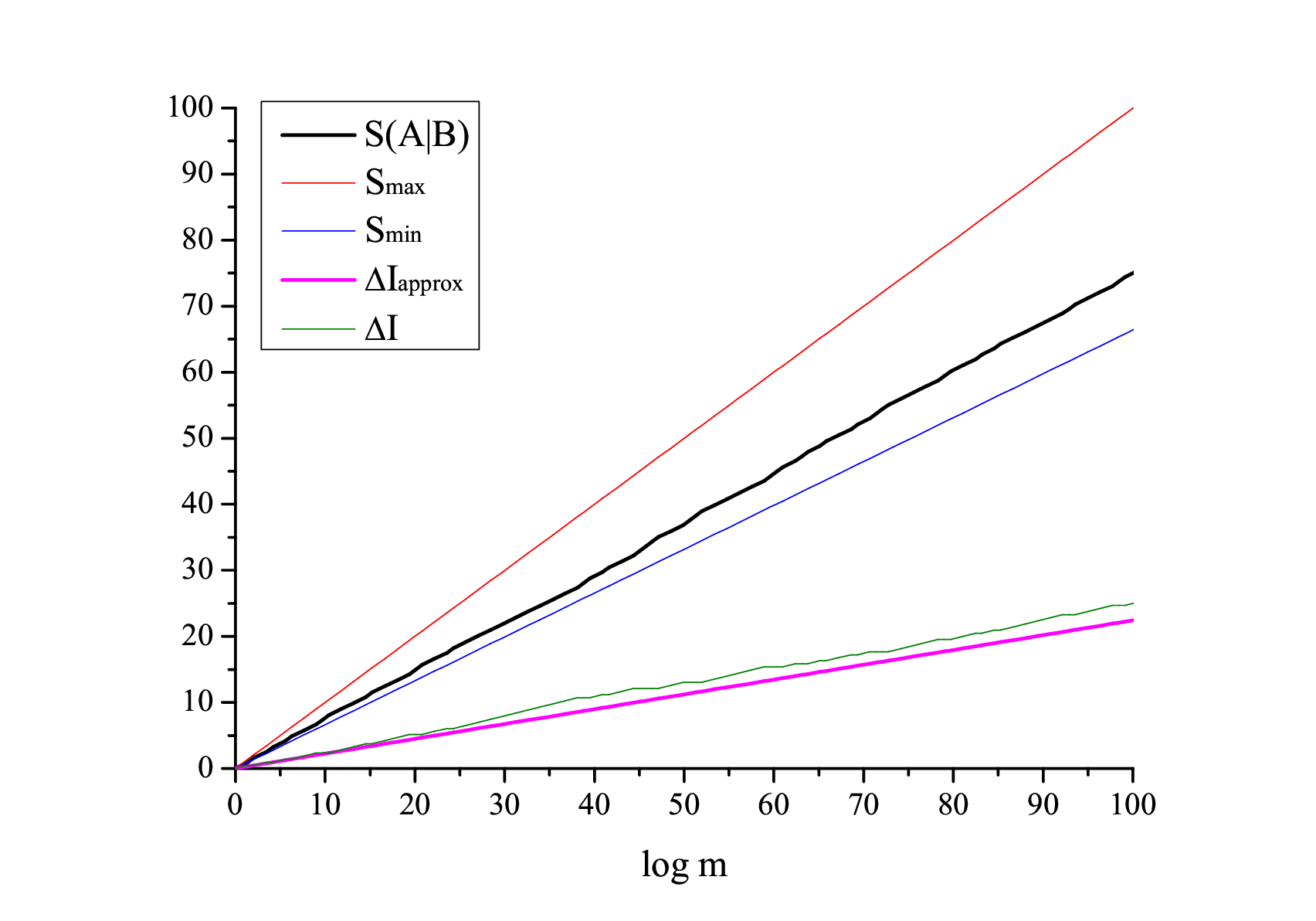}
\includegraphics[scale=0.27]{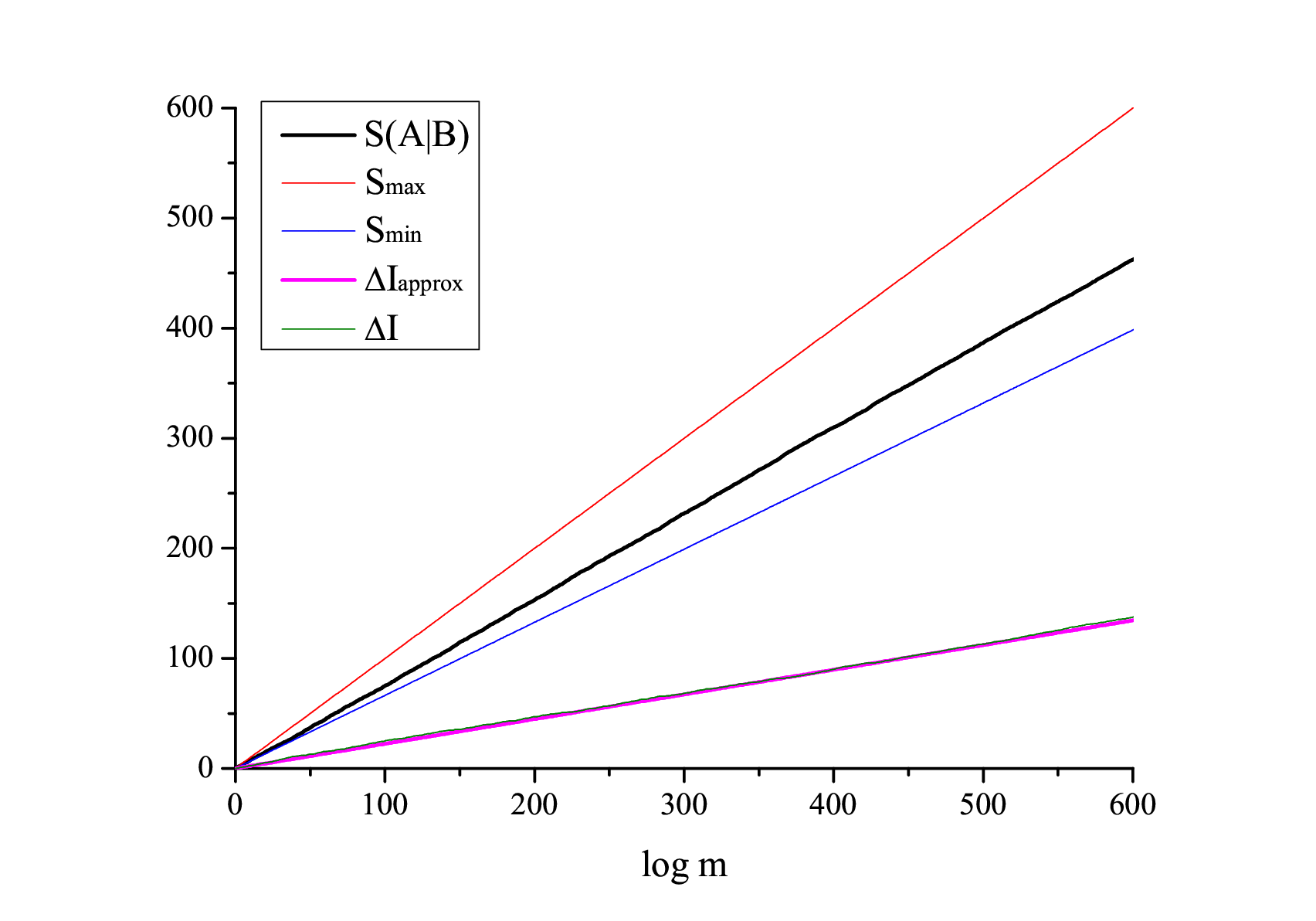}
\includegraphics[scale=0.27]{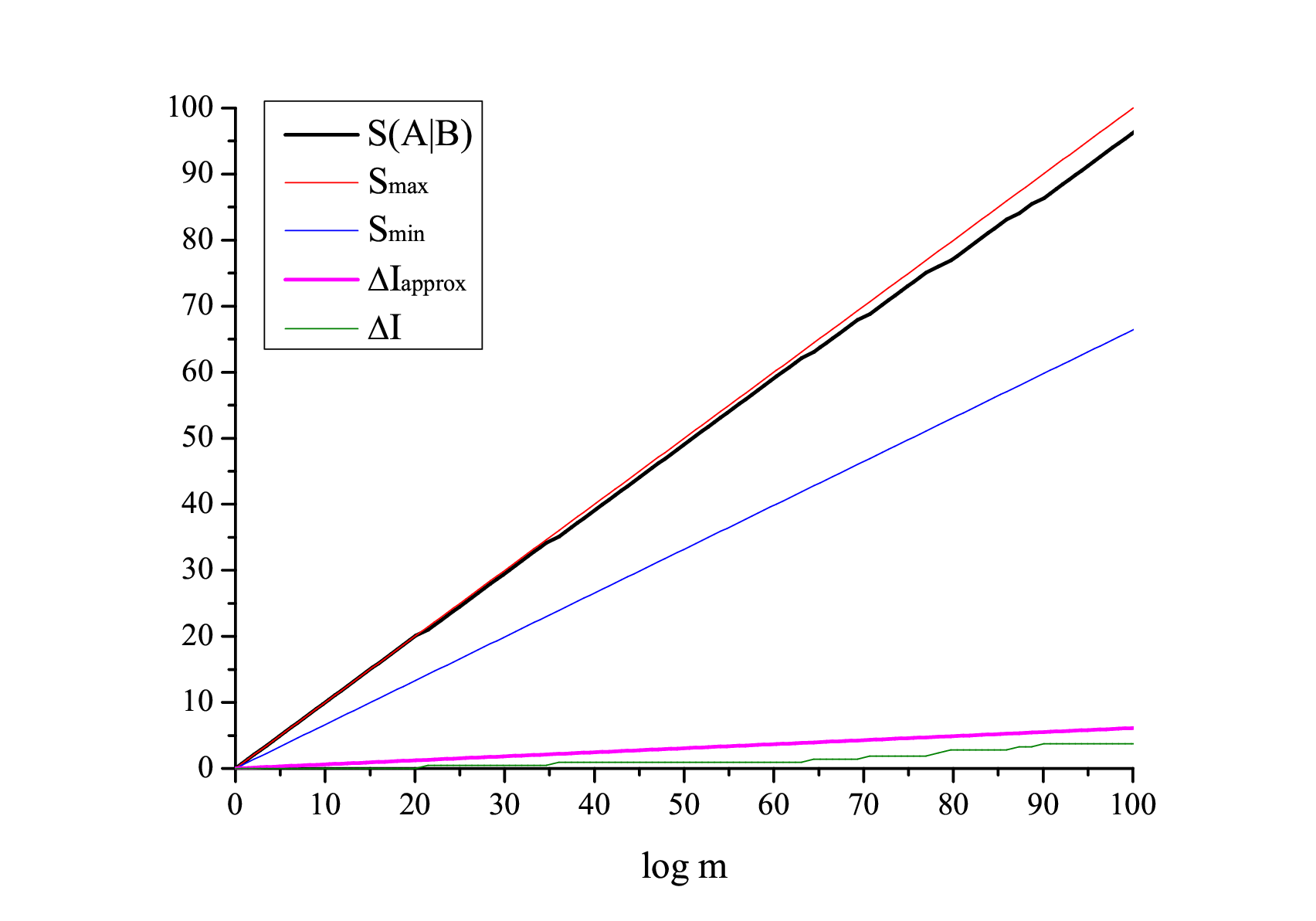}
\includegraphics[scale=0.27]{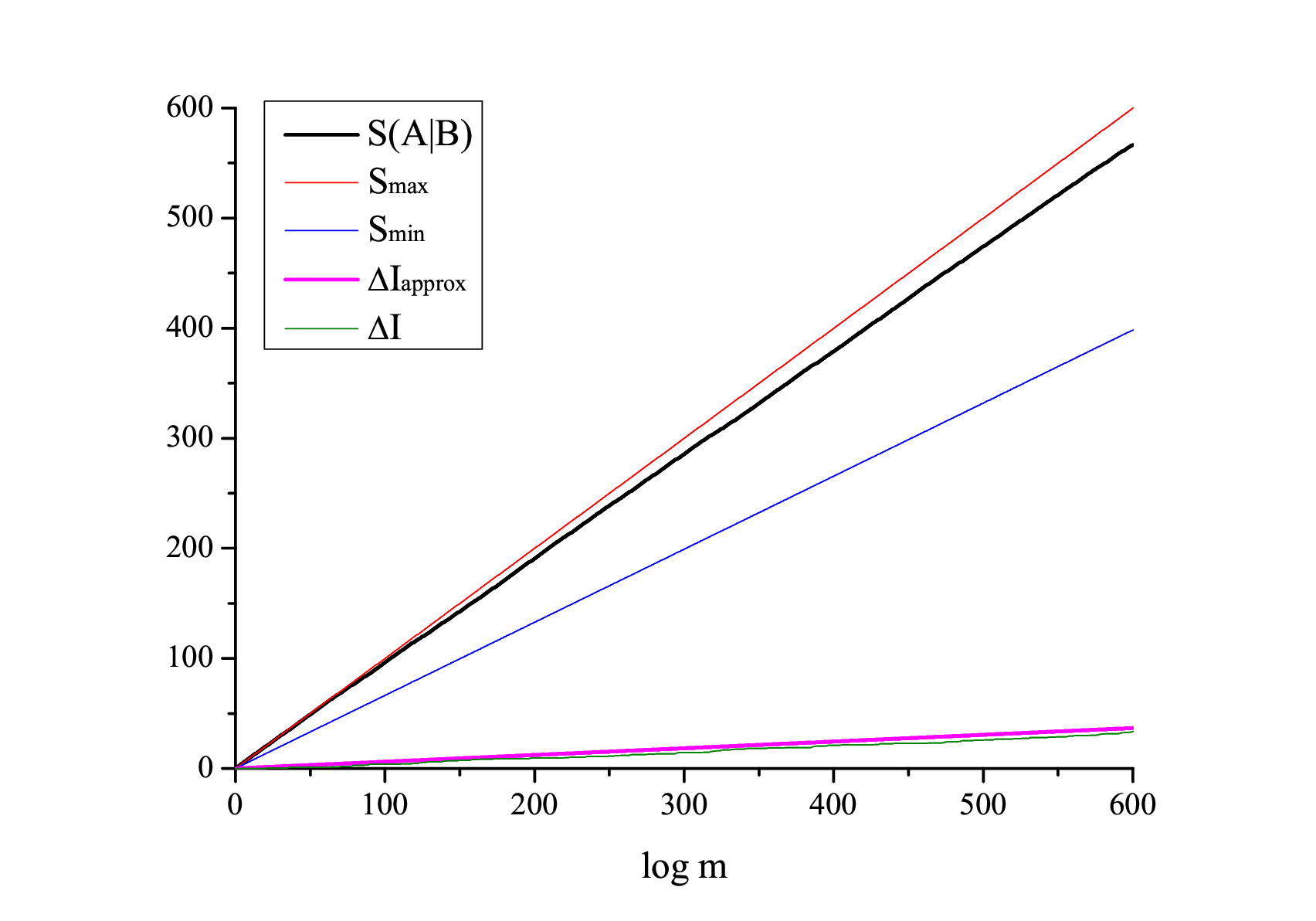}
\caption{\label{fig:ther}Entropy emissions for $p_1 = 0.5$ (top) and $0.9$ (bottom), where $\mathcal{D}\simeq0.92$. Black curves are $S(A|B)$, red and blue lines are $S_{\mathrm{max}}$ (if all processes are maximally entangled) and $S_{\mathrm{min}}$ (if all processes are four-particle creations), respectively, pink lines are analytic fitting curves following Eq.~(\ref{eq:fit}), and green curves are numerical results of $\Delta I$.}
\end{center}
\end{figure}

Indeed, as the number of particles increases, the standard deviation decreases and the entropy reaches a maximum. This means that if maximally entangled pairs interact with many particles, then the entanglement entropy of the half of the total system will almost be the maximum possible value. Alternatively, if maximally entangled particles interact only with a small number of particles, the deviation from the maximum entanglement entropy cannot be negligible. Our results concerning the expectation values of entanglement entropy are consistent with analytic computations by \cite{Bianchi:2019stn}; to be sure, the standard deviations by \cite{Bianchi:2019stn} are lightly different because the analytic form is limited for cases with a large number of particles.

Using this numerical observation, we can now estimate the bias from the originally expected Page curve before the Page time. This is approximated as follows: two processes exist, where one is the creation of a maximally entangled pair and the other is the creation of randomly mixed four-particles. The probability of the first process is $p_{1}$ and that of the second process is $p_{2}$, where $p_{1} \gg p_{2}$ and $p_{1} + p_{2} = 1$. From the first process, for each step, the number of radiation states $\log m$ increases by $\log 2$; simultaneously, the entanglement entropy inside and outside also increases by $\log 2$. Therefore, if $p_{2} = 0$, then $\log m - S \simeq 0$ and the original Page curve is confirmed. However, from the second process, for each step, the number of radiation states increases by $2 \log 2$, because two particles are emitted; the entanglement entropy increases by $\mathcal{D} < 2 \log 2$, where the numerical value of $\mathcal{D}$ is $0.9222$ from the previous simulation. By changing $p_{2}$, we can see the bias from the original Page curve (Fig.~\ref{fig:ther}).

From the numerical simulations (Fig.~\ref{fig:ther}), it is evident that the bias from the original Page curve is indeed proportional to $\log m$. If the total number of events is $\mathcal{N}$, the number of events of the two-particle process is ${\mathcal{N}}_{1} = p_{1} \mathcal{N}$, whereas that of the four-particle process is ${\mathcal{N}}_{2} = p_{2} \mathcal{N}$. For each process of ${\mathcal{N}}_{1}$, the entanglement entropy increases by $\log 2$; for each process of ${\mathcal{N}}_{2}$, the entanglement entropy increases by $\mathcal{D}$. The maximum value of the increase in entanglement entropy is $\log m = {\mathcal{N}}_{1} \log 2 + 2 {\mathcal{N}}_{2} \log 2 = \mathcal{N} (p_{1} + 2 p_{2}) \log 2$. Alternatively, the true entanglement entropy change is ${\mathcal{N}}_{1} \log 2 + {\mathcal{N}}_{2} \mathcal{D} = \mathcal{N} (p_{1} \log 2 + p_{2} \mathcal{D})$. Therefore, the expected emission of information is
\begin{eqnarray}\label{eq:fit}
\Delta I \equiv \log m - S = \left(2 \log 2 - \mathcal{D} \right) \mathcal{N} p_{2} = \left( 2 \log 2 - \mathcal{D} \right) p_{2} \frac{\log m}{(p_{1} + 2 p_{2}) \log 2}.
\end{eqnarray}
This expectation is confirmed by comparing pink lines (analytic estimation) and green curves (numerical computation) in Fig.~\ref{fig:ther}. If $p_{2}$ is too small, then the bias from the Page curve is negligible. However, if $p_{2}$ is not zero and $\log m$ is sufficiently large (i.e., of the order of $\log m \sim p_{2}^{-1}$), there should be a detectable bias.

\section{\label{sec:pos}Possible alternatives}

From the previous discussion, we can conclude that a bias exists from the Page curve if pair-created particles are not maximally entangled with their counterparts. Even though the probability is not high, if a black hole size is large enough, then the accumulated bias will not be negligible. This implies that the bias from the exact thermal state before the Page time becomes
\begin{eqnarray}
\log m - S > 0.
\end{eqnarray}
Accordingly, there must exist a distinguishable effect from Hawking radiation.

Then what is the physical meaning of this conclusion? In this section, we illustrate possible interpretations and consequences.

\begin{figure}
\begin{center}
\includegraphics[scale=0.8]{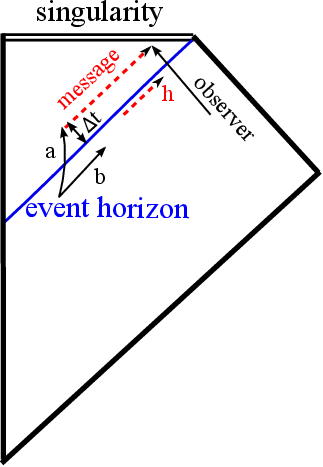}
\caption{\label{fig:Schwarzschild_duplication}A conceptual description of the duplication experiment.}
\end{center}
\end{figure}

\subsection{\label{sec:pos1}Information emission before Page time}

The possible first interpretation is that the relation $\log m - S > 0$ implies that one can read information from Hawking radiation before the Page time. Accordingly, the next natural question is as follows: what is the origin of the emitted information? The first logical guess is that the emitted information concerns collapsed matter. If this is true, then what follows?

We recall the thought experiments of the original version of black hole complementarity \cite{Susskind:1993mu,Hong:2008ga}. As we maintain assumptions A1-A5, the natural consequence is that the information is duplicated, where one copy is inside the horizon (attached by the collapsed matter) and the other copy is outside the horizon (attached by Hawking radiation). One can illustrate this as a simple thought experiment (Fig.~\ref{fig:Schwarzschild_duplication}) \cite{Susskind:1993mu}. An entangled pair of particles ($a$ and $b$) is created and $a$ falls into the black hole. After $a$ falls into the black hole, the information about $a$ is sent to the outgoing direction as a message (red-dashed arrow). Meanwhile, the information about $a$ is emitted by $h$ as Hawking radiation. The observer (black arrow) measures $h$ and $a$; if the message was sent between time $\Delta t$, then the observer can see both $h$ and $a$ before touching the singularity, where $\Delta t$ can be estimated as follows:
\begin{eqnarray}
\Delta t \sim \exp{-\frac{\tau}{M}},
\end{eqnarray}
where $\tau$ denotes the time when $h$ was emitted and $M$ denotes the mass of the black hole. Finally, by comparing with $b$, the observer can notice that $a$ and $h$ are indeed duplicated. If a signal could be sent with some information, it should satisfy the uncertainty relation $\Delta E \Delta t \sim 1$. Hence, the required energy is approximately
\begin{eqnarray}
\Delta E \sim \exp{+\frac{\tau}{M}}.
\end{eqnarray}
If $\Delta E > M$, then this thought experiment is impossible because the necessary energy is larger than the mass of the black hole itself \cite{Susskind:1993mu}. Alternatively, if $\Delta E$ is sufficiently smaller than $M$, then there is no principle to prevent the duplication experiment; hence, the inconsistency of assumptions A1-A5 is revealed. Then, after a simple computation, we can check that if the information is emitted before the scrambling time \cite{Hayden:2007cs}, $\sim M \log M$, an observer who sees a duplication of information can exist, which violates the principle of quantum mechanics.

In our case, to see a bit of information (i.e., $\Delta I \sim \mathcal{O} (1)$), the condition is
\begin{eqnarray}\label{eq:cond1}
\log m_{\mathrm{f}} \sim \frac{1}{p_{2}},
\end{eqnarray}
where $m_{\mathrm{f}}$ denotes the required radiation degree of freedom to have $\Delta I \sim \mathcal{O} (1)$. Hence, as the probability of a randomly mixed four-particle process ($p_{2}$) decreases, we must wait for longer periods of time. Note that $\log m_{\mathrm{f}}$ is related to the thermal entropy of the radiation:
\begin{eqnarray}\label{eq:cond2}
\log m_{\mathrm{f}} = \frac{A_{\mathrm{i}} - A_{\mathrm{f}}}{4} = \pi M_{\mathrm{i}}^{2} \left( 1 - \frac{M_{\mathrm{f}}^{2}}{M_{\mathrm{i}}^{2}} \right),
\end{eqnarray}
where $A_{\mathrm{i,f}}$ denotes the area of the horizon and $M_{\mathrm{i,f}}$ denotes the mass of the black hole at initial and final states. For simplification, let us define the ratio $M_{\mathrm{f}}/M_{\mathrm{i}} \equiv x$, which is less than one.

Following the Stefan-Boltzmann law, the required time to see such a change in the mass is proportional to $M^{3}$; i.e.,
\begin{eqnarray}
\tau \sim M_{\mathrm{i}}^{3} \left( 1 - x^{3} \right).
\end{eqnarray}
Therefore, by comparing Eqs.~(\ref{eq:cond1}) and (\ref{eq:cond2}), we obtain
\begin{eqnarray}
\tau \sim \frac{M_{\mathrm{i}}}{p_{2}} \left( \frac{1-x^{3}}{1-x^{2}} \right).
\end{eqnarray}
If this has the same order of the scrambling time, we obtain two conditions:
\begin{eqnarray}
M_{\mathrm{i}}^{2} &\sim& \frac{\log M_{\mathrm{i}}}{1-x^{3}}, \\
\frac{1}{p_{2}} \left( \frac{1-x^{3}}{1-x^{2}} \right) &\sim& \log M_{\mathrm{i}}.
\end{eqnarray}
Therefore, $x \simeq 1$ is required and at the same time
\begin{eqnarray}
p_{2} \sim \frac{1}{\log M_{\mathrm{i}}}
\end{eqnarray}
is expected. Therefore, even though $p_{2}$ is small enough, if $p_{2}$ is not zero, then $x$ and $M_{\mathrm{i}}$ exist, which allows for the duplication experiment. This is not surprising because if $x$ is small but $M$ is large, then the emitted number of states is also large, even though the time scale is relatively short (compared with the scrambling time).

Therefore, if the emitted information concerns the collapsed matter, then one can see the duplication of information. Of course, this conclusion is not acceptable. Therefore, the reasonable interpretation is that \textit{if pair-created particles are not maximally entangled with their counterparts, then assumptions A1-A5 are not consistent with each other, even before Page time.}

\subsection{\label{sec:pos2}Firewalls or non-local effects before the Page time}

Another possible interpretation is to consider the violation of A2 or A3, i.e., applying non-local interactions \cite{Giddings:2012gc} or the firewall \cite{Almheiri:2012rt} (or any other kind of membrane \cite{Thorne:1986iy}) before the Page time. Accordingly, whether such membranes are required or whether the natural laws are violated before the Page time needs to be examined.

If a firewall (or any other kind of membrane) exists, its effect is naked in principle \cite{Hwang:2012nn}. If any effects are naked after the Page time, then there would not be any astrophysical evidence of them; however, if any effects are naked before Page time, then evidence of them can be obtained by astrophysical experiments, e.g., from gravitational wave observations. Recently, membranes approximal to the event horizon have been examined using gravitational wave data \cite{Abedi:2016hgu}. Indeed, if the requisite evidence cannot be obtained from gravitational wave experiments, then we can surely say that firewalls (or any other kind of membrane) are not the candidate solution to the information loss paradox. Accordingly, we can say that astrophysical observations can confirm non-local realizations of quantum gravity if the non-local effects happen before the Page time. We leave this topic for future investigations.

\subsection{\label{sec:pos3}Formation of a monster}

If we maintain A1, A2, and A3, then one may guess that A4 is violated. To explain how the violation of A4 can solve the inconsistency, we must discuss the information flow of a bipartite system \cite{Alonso-Serrano:2015bcr}.

Let us define again that there is a bipartite system, where $A$ has the number of states, $m$, and the complement part, $B$, has the number of states, $n$. The total number of states is $\Omega = n \times m$. Accordingly, we can define three kinds of information ($S(A|B)$ is the entanglement entropy between $A$ and $B$):
\begin{itemize}
\item[--] Information of $A$: $I_{A} \equiv \log m - S(A|B)$.
\item[--] Information of $B$: $I_{B} \equiv \log n - S(B|A)$.
\item[--] Mutual information between $A$ and $B$: $I_{AB} \equiv S(A|B) + S(B|A) - S(A \cup B)$.
\end{itemize}
If we assume that the total system is in a pure state, then $S(A|B) = S(B|A)$ and $S(A\cup B) = 0$. Therefore,
\begin{eqnarray}
I_{A} + I_{B} + I_{AB} = \log nm = \log \Omega.
\end{eqnarray}
If the total number of states, $\Omega$, is a constant, then the sum of three functions is a constant and accordingly, the total information is preserved. However, even though $\Omega$ varies with time, total system can still be in a pure state, and the evolution can still be unitary \cite{Page:2013dx}.

\begin{figure}
\begin{center}
\includegraphics[scale=0.7]{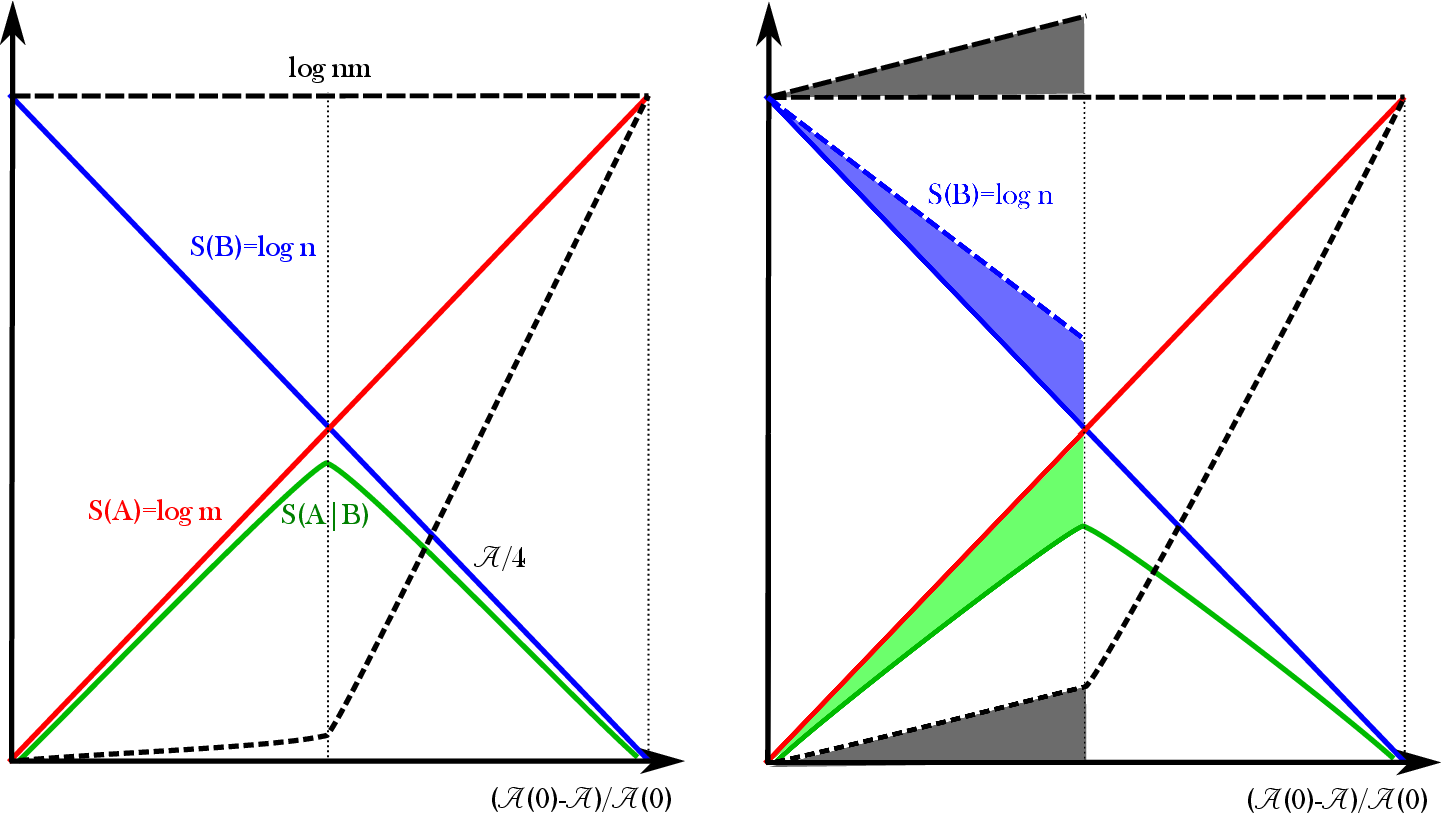}
\caption{\label{fig:newpage}Left: Traditional Page curve, where the black dashed line is the total number of states, the blue line is the black hole entropy, the red line is the radiation entropy, the green curve is the entanglement entropy, and the black dotted curve is the information of radiation $I_{A}$. Right: When the total number of states increases (gray colored region of the top).}
\end{center}
\end{figure}

In this regard, even though $I_{A} = \Delta I$ is greater than zero, if the total number of states is increased by the same factor as $\Delta I$, then $I_{A}$ does not concern the transferred information from $B$; rather, this would just be an artifact due to the increase in the number of states (Fig.~\ref{fig:newpage}).

The left of Fig.~\ref{fig:newpage} is the traditional Page curve and the right of Fig.~\ref{fig:newpage} is the case when the total number of states increases (gray colored region of the top). In this figure, the blue and red lines are meaningful because they represent the areal entropy change of a black hole; in terms of the semi-classical description, the sum of the thermal entropy of radiation and the areal entropy will be a constant. Accordingly, since the total number of states increases, the same amount of the Boltzmann entropy must be accumulated inside (blue colored region), which becomes greater than the areal entropy (blue line). As a result, even though large differences exist between the radiation entropy and the entanglement entropy (green colored region), the radiation information may not concern the collapsed matter (gray colored region of the bottom) but, rather, constitutes an effect caused by the increase in the total number of states.

This apparently resolves a paradox, but it also requires that $S(B) = \log n$ is greater than the Bekenstein-Hawking entropy; a \textit{monster} is obtained. Such a difference in radiation information is negligible if the black hole mass is large enough. However, the same process can be repeated many times. Let us consider the case where a black hole evaporates from mass $M$ to $M'$, and that the statistical entropy is accumulated by $\Delta I$. Here, the amount of $\Delta I$ in itself might be small. However, by adding more matter to the black hole, we can increase the mass of the black hole from $M'$ to $M$. As we repeat the evaporation from $M$ to $M'$, the entropy accumulation will be $2 \Delta I$. In principle, we can repeat this many times, which means that the statistical entropy will unboundedly be larger than the Bekenstein-Hawking entropy. Accordingly, we can conclude that \textit{once we allow a monster, the number of states inside a black hole can be unbounded in principle.}

Of course, we will suffer from the traditional problems of the remnant picture \cite{Giddings:1994qt,Chen:2014jwq}, which means that the number of states may become unlimitedly large even though the areal size is finite. Then, it will bring difficult theoretical problems once again.

\section{\label{sec:com}Comments on recent developments}

We have considered three alternative possibilities. First, we considered the possibility that information is carried out before the Page time via Hawking radiation. However, this results inconsistency; information might be duplicated. Second, we considered firewalls or non-local effects. Third, we considered remnants or monster-like object formation, where the Boltzmann entropy exceeds its areal entropy. 

Regarding the second possibility, we need a mechanism to realize non-local interactions. In this context, it is worthwhile to compare our results with the recent discussion in the string community about the entanglement entropy issue. Based on the quantum extremal surface technique \cite{Engelhardt:2014gca}, one could obtain the entanglement entropy of an evaporating black hole \cite{Almheiri:2019hni}. According to their computation, before the Page time, the entanglement entropy monotonically increases because the dominant contribution comes from the original Hawking's saddle, while after the Page time, the dominant contribution comes from a new saddle that can be interpreted as the replica wormholes \cite{Almheiri:2019qdq}. However, these saddles will contribute after the Page time, and it is hard to believe that they can contribute in the early stage of black hole evaporations.

In addition to this, still it is unclear to understand the detailed process to carry out information from the black hole. The difficulty comes from the fact that the entanglement entropy computation is a path integral of the density matrices, while the authentic quantum mechanics is about a path integral of the quantum states \cite{Chen:2021jzx}. Because of this problem, even though one computes the entanglement entropy on the Euclidean geometry, it is very unnatural to analytically continue to the Lorentzian signatures. There have been some trials to interpret in the Lorentzian signatures \cite{Marolf:2020rpm}, but it is still too premature to conclude the final interpretation.

Probably, the third option might be the most conservative approach, but we need more justifications. In this regards, although it is a minor opinion, some researchers proposed an idea that if we do not keep the central dogma \cite{Buoninfante:2021ijy}, i.e., the assumption A4 (Bekenstein-Hawking entropy as statistical entropy), one may overcome the difficulties of the information loss paradox. As proposed in \cite{Buoninfante:2021ijy}, it is conceptually possible to derive a monster-like state that has more statistical entropy than its areal entropy. Recent computations from Euclidean path integral approach \cite{Chen:2021jzx} also indicates the same direction. Our paper proposes a new mechanism that can derive a monster-like state; the interesting new point of ours is that the monster-like state can be derived even before the Page time.

Therefore, although it is not conclusive, our approach sheds some interesting intuitions to the recent discussion of the string community. For example, if the monster-like state construction is allowed via a matter field with self-interactions as we expected in our paper, the contribution from the islands can be effective even before the Page time, because usually one believes that the monster-like behavior is one signature of the existence of the island contribution, although this is not quite consistent to the so-called mainstream picture of the information loss paradox. Further discussion is beyond the scope of this paper; we left them for future research topics.

\section{\label{sec:con}Conclusion}

In this paper, we investigated the particle emission process from a black hole using spin-1/2 toy models. It is worthwhile to mention that our analysis have some limitations. We considered spin-1/2 particles with finite degrees of freedom, while quantum fields have infinite degrees of freedom. However, our analysis will shed some lights to the information loss paradox, because we regard that the quantum state has a finite entropy as well as a finite degrees of freedom. Therefore, in terms of qubits and their entanglements, the same relationship must be applicable not only for particles but also for quantum states of fields.

In doing so, the traditional consensus was that, before the Page time, Hawking radiation is generated by a particle-antiparticle pair creation, where two particles are maximally entangled and form a separable state. Alternatively, we considered a situation where two maximally entangled pairs interact with each other; or where a separable four-particle system is created from a vacuum (Sec.~\ref{sec:pre}). Accordingly, it was assumed that the infalling two particles are randomized with the outgoing two particles, which implies that the incoming particles do not satisfy the maximum entanglement condition with the outgoing particles.

To the authors' knowledge, there is no fundamental reason to abandon this possibility, although the probability can be less dominant. As time passes, the bias from the original Page curve before the Page time is accumulated (Sec.~\ref{sec:bia}).

Such a bias can be interpreted as the Hawking radiation not being in the thermal state, and hence, one can obtain information from Hawking radiation. However, from a simple thought experiment (Sec.~\ref{sec:pos1}), we can prove that this interpretation results in a fatal inconsistency. 

If it is not interpreted that the bias from the thermal state does not imply the emission of information of the collapsed matter, then either semi-classical quantum field theory is incorrect (Sec.~\ref{sec:pos2}) or else additional states are accumulated inside the horizon (Sec.~\ref{sec:pos3}); i.e., a monster is formed. If the monster forms before the Page time, then it is possible to imagine that an unbounded number of states are accumulated inside a black hole, which is theoretically unintuitive. 

If a firewall (or non-local interactions) appears before the Page time, it is probable that such a radically new effect can be confirmed or falsified by future astrophysical observations. At the first glimpse, it is fascinating that astrophysical observations can test or falsify some hypothesis of quantum gravity. However, this conclusion also implies that the information loss paradox is indeed worsened, since the paradox appears not only after but also before the Page time. 

If our three alternatives are not satisfactory, then the remaining alternative is that information is lost or, at least, effectively lost \cite{Lee:2015rwa}. We need to know which one is the correct idea. It will be useful to understand the information transfer process not only after but also before the Page time, based on field theory or quantum gravity. Of course, this was beyond the scope of this paper and it was, therefore, left for future research.

\section*{Acknowledgment}

DY was supported by a 2-Year Research Grant of Pusan National University. JMB and HZ were supported by the Daegu Gyeongbuk Institute of Science and Technology (DGIST) Undergraduate Group Research Project (UGRP) grant.

\newpage

\section*{Appendix A. Mathematical prescriptions of spin-1/2 systems}

Quantum entanglements of a system are generally measured by the entanglement entropy of the system. There are many different definitions of entanglement entropy, one of which is the von Neumann entropy. For a system with the density matrix, $\rho$, the von Neumann entropy is defined by $S(\rho)=-\operatorname{Tr}{\rho \log \rho}$. The log of a matrix can be calculated after diagonalizing the matrix, i.e.,
\begin{eqnarray}
\rho \rightarrow \operatorname{U^{-1}} \rho \operatorname{U}=D=
\begin{bmatrix}
\lambda_{1} & & \\
& \ddots & \\
& & \lambda_{r}
\end{bmatrix}.
\end{eqnarray}
Then,
\begin{eqnarray}
S(\rho)=-\operatorname{Tr}{\rho \log \rho}=-\sum_{i=1}^n\lambda_i \log \lambda_i.
\end{eqnarray}
To calculate the entanglement entropy of a subsystem, we take partial trace of the density matrix of the total system to obtain the density matrix of the subsystem. The total density matrix is, say, $\rho=\rho_A \bigotimes \rho_B$; we are interested in the entanglement entropy of $\rho_A$. Then,
\begin{eqnarray}
\rho_A=\operatorname{Tr_B}{\rho},\quad \text{where}\quad  \rho{_A}_{[i,j]}=\sum_{k=1}^{\text{dim}(\rho_B)} \rho_{[i,j]\bigotimes[k,k]},
\end{eqnarray}
and the entanglement entropy of $\rho_A$ can be calculated same as above,
$S(A|B)=-\operatorname{Tr}{\rho_A \log(\rho_A)}$.

In this paper, we considered spin-1/2 systems. For a given $n$-particle system, the state can be expressed by 
$|{\Psi}_{\textrm{total}} \rangle = \sum_{i_1...i_n} c_{i_1...i_n}|i_1...i_n\rangle, (\text{$i$=1 or 2, where 1 denotes spin up and 2 denotes spin down}).$ The density matrix of the total system is given by
$\rho=|\Psi\rangle \langle\Psi|=\sum_{i_1...i_n, j_1...j_n}c_{i_1...i_n}c_{j_1...j_n}^{*}|i_1...i_n\rangle \langle j_1...j_n|.$
Then the entanglement entropy of this system can be calculated by following the process described above.  

\section*{Appendix B. Details on the swap operation}

To approximately describe the randomizing process, in Fig.~\ref{fig:randomizing}, we introduced the swap operation. In doing so, we consider two pairs of maximally entangled particles, namely, $\rho_{a\cup a'}$ and $\rho_{b\cup b'}$. Therefore, the total number of particles is four.

In the spin-1/2 particle system, we can define
\begin{equation}
\left| \uparrow \right> = \begin{pmatrix} 1 \\ 0 \end{pmatrix}, \;\;\;\;\;\; \left| \downarrow \right> = \begin{pmatrix} 0 \\ 1 \end{pmatrix}.
\end{equation}
Using the above definitions, we can create two maximally entangled states and their density matrices, $\rho_{a\cup a'}$ and $\rho_{b\cup b'}$, where
\begin{eqnarray}
\rho_{a\cup a'} &=& \frac{1}{\sqrt{2}} \left( \left| \uparrow \downarrow \right> \pm \left| \downarrow \uparrow \right> \right),\\
\rho_{b\cup b'} &=& \frac{1}{\sqrt{2}} \left( \left| \uparrow \downarrow \right> \pm \left| \downarrow \uparrow \right> \right).
\end{eqnarray}
Hence, the total density matrix becomes
\begin{equation}
\rho = \rho_{a\cup a'} \otimes \rho_{b\cup b'}.
\end{equation}
With regard to the signs of the spin combinations, there are four possible combinations of the total density matrices; however, there are no significant differences in selecting any combination.

Then, we need to define the swap operator: 
\begin{equation}
\mathcal{O}^{(i^* , j^*)} = SW_{\rho}^{i^* : j^*} \otimes \mathcal{I},
\end{equation}
where $\mathcal{I}$ is an $2^4 \times 2^4$ identity matrix. For $i^*, j^* \in \left[ i_1,i_{2^4}\right]$, the elements of $SW_\rho$ become 
\begin{eqnarray}
SW_{\rho}^{i^* : j^*}[i,j] &=& \delta_{ij}\qquad \;\;\;\;\;\;\;\;\;\;\; \textrm{if}\quad \, i,j \neq i^*,j^*,\\
%\end{equation}
%\begin{equation}
SW_{\rho}^{i^* : j^*}[i^*,i^*] &=& SW_{\rho}^{i^* : j^*}[j^*,j^*] = 0,\\
%\end{equation}
%\begin{equation}
SW_{\rho}^{i^* : j^*}[i^*,j^*] &=& SW_{\rho}^{i^* : j^*}[j^*,i^*] = 1, 
\end{eqnarray}
The physical meaning of this operation is to select and exchange an arbitrary row or column ($i^*$ and $j^*$, respectively). Since we have defined a swap operator, we can use it to mix our total density matrix, $\rho$.

\end{document}